\documentclass[9pt,twocolumn,twoside]{osajnl}

\journal{ol}

\setboolean{shortarticle}{true}

		\catcode`@=11
		\def\nvphantom{\v@true\h@false\nph@nt}
		\def\nhphantom{\v@false\h@true\nph@nt}
		\MakeRobust{\vphantom}
		\def\nphantom{\v@true\h@true\nph@nt}
		\def\nph@nt{\ifmmode\def\next{\mathpalette\nmathph@nt}\else\let\next\nmakeph@nt\fi\next}
		\def\nmakeph@nt#1{\setbox\z@\hbox{#1}\nfinph@nt}
		\def\nmathph@nt#1#2{\setbox\z@\hbox{$\m@th#1{#2}$}\nfinph@nt}
		\def\nfinph@nt{\setbox\tw@\null \ifv@ \ht\tw@\ht\z@ \dp\tw@\dp\z@\fi\ifh@ \wd\tw@-\wd\z@\fi \box\tw@}
		\usepackage{stackengine}[2013-10-15]
		\usepackage{scalerel}
		
		\renewcommand{\exp}[1]{\text{exp}({#1})}

		\newcommand{\Real}[1]{\text{Re}\!\left({#1}\right)}

		\newcommand{\integral}[4]{\int_{#1}^{#2}{#3}\ifthenelse{\isempty{#4}}{}{\,\text{d}{#4}}}
		\newcommand{\ointegral}[3]{\oint_{#1}{#2}\ifthenelse{\isempty{#3}}{}{\,\text{d}{#3}}}
		\usepackage{scalerel}[2016-12-29]

		\usepackage[makeroom]{cancel}
		
		\renewcommand{\geq}{\geqslant}
		
		\newcommand*{\eq}[1]{\begin{eqnarray}#1\end{eqnarray}}
		
		\usepackage{etoolbox}
		\AtBeginEnvironment{eqnarray}{\setlength{\arraycolsep}{2pt}}

		\newcommand{\figwidthc}{0.88\linewidth}

		\usepackage{pgfplots}
		\pgfplotsset{compat=1.9}
		\usepgfplotslibrary{fillbetween}
		\definecolor{blood}{rgb}{0.8,0,0}
		\definecolor{sodium}{rgb}{1,0.75,0}
		\definecolor{chlorophyll}{rgb}{0,0.8,0}
		\definecolor{iodine}{rgb}{0.6,0.2,0.9}
		\usepackage{subcaption}

		\usepackage{xifthen}
		\newcommand{\article}[6]{#1 (#2): \textit{#3}, #4 #5, \mbox{#6}}

		\newcommand{\book}[6]{#1 (#2): \textit{#3}, \ifthenelse{\isempty{#4}}{}{\ifthenelse{\equal{\detokenize{#4}}{\detokenize{2}}}{second}{\ifthenelse{\equal{\detokenize{#4}}{\detokenize{3}}}{third}{\ifthenelse{\equal{\detokenize{#4}}{\detokenize{4}}}{fourth}{\ifthenelse{\equal{\detokenize{#4}}{\detokenize{5}}}{fifth}{\ifthenelse{\equal{\detokenize{#4}}{\detokenize{6}}}{sixth}{\ifthenelse{\equal{\detokenize{#4}}{\detokenize{7}}}{seventh}{\ifthenelse{\equal{\detokenize{#4}}{\detokenize{8}}}{eighth}{\ifthenelse{\equal{\detokenize{#4}}{\detokenize{9}}}{ninth}{#4}}}}}}}} edition, }#5\ifthenelse{\isempty{#6}}{}{, \mbox{#6}}}
		\newcommand{\bookarticle}[7]{#1 (#2): \textit{#3}, in \textit{#4}\ifthenelse{\isempty{#5}}{}{ #5}, #6, \mbox{#7}}

		\newcommand{\Mg}{M_\text{g}}
		
		\newcommand{\nb}{n_\text{b}}
		\renewcommand{\ng}{n_\text{g}}
		\newcommand{\nw}{n_\text{w}}
		\newcommand{\thetab}{\theta_\text{b}}
		\newcommand{\thetai}{\theta_\text{i}}
		\newcommand{\thetat}{\theta_\text{t}}
		\newcommand{\thetaw}{\theta_\text{w}}

\title{Detecting and analysing wavelength-scale optical gradients at an interface by their effects on the internal reflectance near the critical angle}

\author[1]{Omar V\'azquez-Estrada}
\author[2]{Anays Acevedo-Barrera}
\author[2,*]{Alexander Nahmad-Rohen}
\author[2]{Augusto Garc\'{\i}a-Valenzuela}

\affil[1]{Tecnol\'ogico Nacional de M\'exico / ITS de Tantoyuca, Desviaci\'on Lindero Tametate S/N, La Morita, Tantoyuca, Veracruz, M\'exico}
\affil[2]{Instituto de Ciencias Aplicadas y Tecnolog\'{\i}a, Universidad Nacional Aut\'onoma de M\'exico, Circuito Exterior S/N, Avenida Universidad 3000, Coyoac\'an, Ciudad de M\'exico, M\'exico}

\affil[*]{Corresponding author: alexander.nahmad@icat.unam.mx}

\begin{abstract}
Light's internal reflectivity near a critical angle is very sensitive to the angle of incidence and the optical properties of the external medium near the interface. Novel applications in biology and medicine of subcritical internal reflection are being pursued. In many practical situations the refractive index of the external medium may vary with respect to its bulk value due to different physical phenomena at surfaces. Thus, there is a pressing need to understand the effects of a refractive-index gradient at a surface for near-critical-angle reflection. In this work we investigate theoretically the reflectivity near the critical angle at an interface with glass assuming the external medium has a continuous depth-dependent refractive index. We present graphs of the internal reflectivity as a function of the angle of incidence, which exhibit the effects of a refractive-index gradient at the interface. We analyse the behaviour of the reflectivity curves before total internal reflection is achieved. Our results provide insight into how one can recognise the existence of a refractive-index gradient at the interface and shed light on the viability of characterising it.
\end{abstract}

\setboolean{displaycopyright}{true}

\begin{document}\sloppy

\maketitle

\section{Introduction}\label{sec-intro}

Total internal reflection (TIR) occurs at a flat interface between two transparent media in an internal-reflection configuration for angles of incidence greater than the critical angle. By internal-reflection configuration it is meant that light is incident from a medium of higher refractive index than that of the ``external'' medium. For angles of incidence slightly smaller than the critical angle, the reflectivity has a steep dependence on the angle of incidence. Ideally, for an incident plane wave, the angular derivative of the reflectance diverges as the angle of incidence approaches the critical angle from the smaller-angle side and is identically zero for larger angles of incidence. Therefore, at the critical angle there is an infinitely large discontinuity in the angular derivative of the reflectance. Locating the critical angle for light reflection in an internal-reflection configuration is thus a convenient and accurate way of measuring the refractive index of the external medium assuming that of the internal medium is known. This is the working principle of Abbe and Abbe-type refractometers \cite{ref-Rheims-MST8,ref-Meeten-MST2,ref-ContrerasTello-MST25,ref-Meeten-MST6,ref-Liu-AO35}. Likewise, the reflectivity of a well-collimated beam (e.g.~a laser beam) near the critical angle in an internal-reflection configuration at a glass interface can be used as a simple and sensitive means of measuring and monitoring the refractive index of the external medium \cite{ref-MarquezIslas-MST31,ref-GarciaValenzuela-OE41}.

TIR microscopy and TIR fluorescence (TIRF) microscopy are established techniques used in physics and biology to track single molecules and record concentration maps of molecules near a glass interface \cite{ref-Liu-PNAS111,ref-Sun-JLA16,ref-ElArawi-OL44,ref-Szalai-NC12}. TIRF microscopy uses light incident from the glass side at a well-defined angle of incidence above the critical angle. Only a thin layer of the external medium is illuminated by an evanescent wave. The penetration depth of light can be controlled with the angle of incidence, allowing depth information to be obtained. Recently, near-total internal reflection microscopy was proposed as a technique able to map the local refractive index inside cells \cite{ref-Bohannon-MM23}. This technique employs the steep increase in the reflectivity with the angle of incidence just before the critical angle to achieve a high sensitivity in refractive-index measurement. However, the reflectivity near the critical angle is very sensitive to the surface conditions at the wavelength scale. A thin layer with a different refractive index between the interior of a cell and the medium of incidence must be considered. Thus, measurement of the cell's refractive index requires simultaneously measuring the thickness and the refractive index of the intervening thin layer.

Some of these works \cite{ref-MarquezIslas-MST31,ref-Bohannon-MM23} demonstrate the practicality and high sensitivity of near-critical-angle reflectometry, leading to exciting new applications.

In view of the aforementioned works, we may ponder the relevance of a refractive-index gradient in the external medium in an internal-reflection configuration. Refractive-index gradients can appear in many physical systems due to concentration or temperature gradients at the interface. Concentration gradients can in turn occur due to flows, sedimentation, diffusion processes, ablation processes, etc. Other systems, such as a disordered monolayer of transparent particles, can be studied as an equivalent graded-refractive-index film \cite{ref-Diamant-JOSAA29}, reproducing the reflectivity predictions of more sophisticated multiple-scattering models for sub-wavelength particle sizes at normal incidence. For reflectivity measurements in biological samples, this simplification in the electromagnetic treatment is applicable by choosing the adequate illumination wavelength, allowing the exploration of the migration of cells from a surface, the exchange of liquids through a cell's membrane or a change in the average cell size. Refractive-index profiles can also be fabricated to create antireflective coatings or wave guides, in which numerical characterisation is important for the optical testing of such devices \cite{ref-Diamant-JOA11}. Thus, in view of the novel techniques based on sub-critical-angle reflectometry, whether one can recognise the presence of a refractive-index gradient and even characterise it by angle-resolved reflectivity measurements appears now as a relevant question worth immediate attention.

We are not aware of much theoretical work aimed at studying internal reflectance from media with spatially variable refractive index or layered media which could help us not only better understand the optical phenomena occurring around the critical angle, but also think about possible applications in biology and synthetic systems. This would lead to a greater understanding of the interaction of electromagnetic waves with biological media or thin films deposited on a substrate, to mention just a few examples \cite{ref-Mangini-URSIISET22,ref-Lekner-JASA87,ref-Culshaw-AO11}.

Here, we study theoretically the reflectivity of light in an internal-reflection configuration with a depth-dependent refractive index in the external medium. We aim to obtain physical insight on the behaviour of the angle dependence of the reflectivity in this problem for use in the future development of TIR microscopy and related techniques. We will perform our analysis assuming an incident plane wave; therefore, since we are considering a flat interface, reflectivity and reflectance are synonymous.

\section{Theory}\label{sec-theory}

\begin{figure}[t!]\begin{center}\begin{subfigure}[t]{\figwidthc}\begin{center}{\includegraphics[width=\figwidthc]{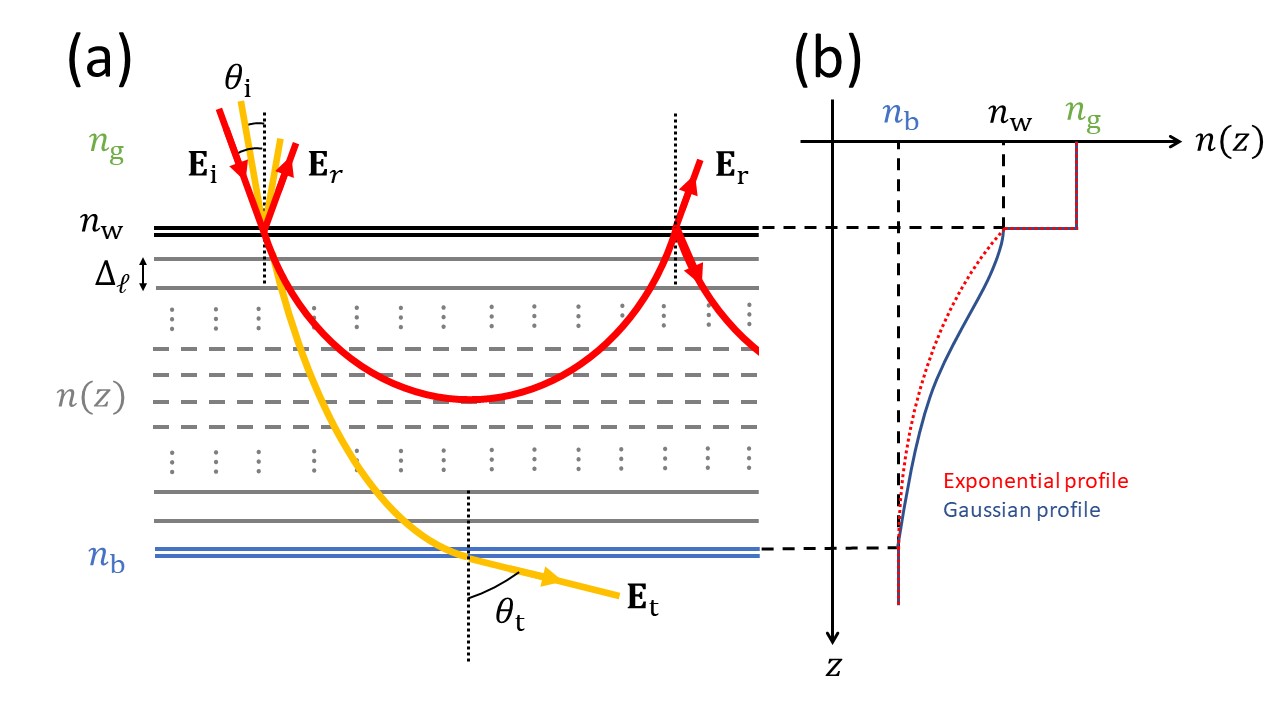}}\end{center}\end{subfigure}\\\begin{subfigure}[t]{\figwidthc}\begin{center}{\includegraphics[width=\figwidthc]{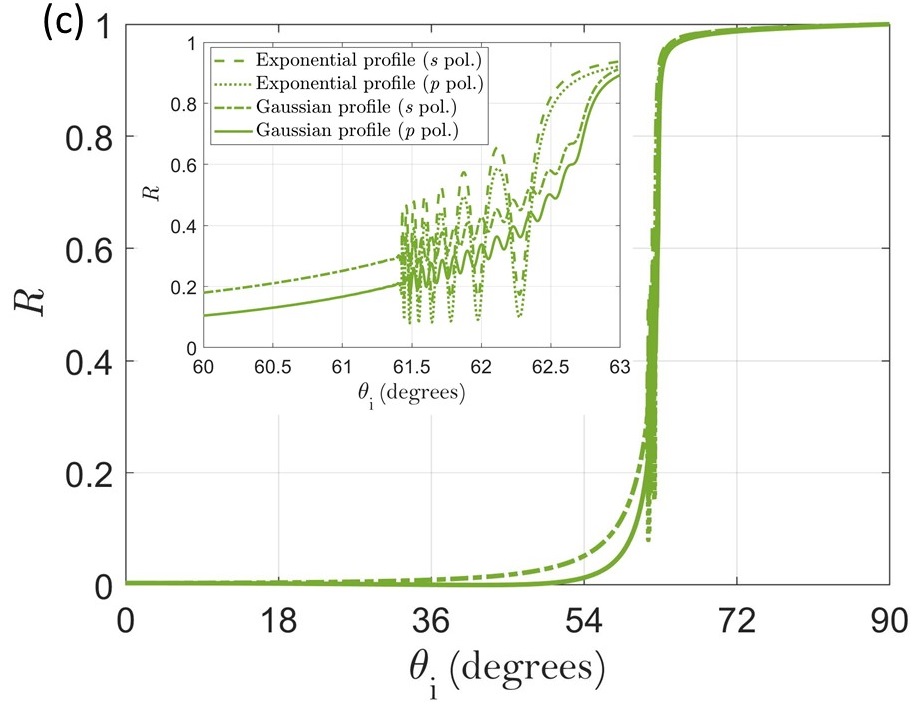}}\end{center}\end{subfigure}\captionsetup{singlelinecheck=off}\end{center}\caption[.]{(a) Schematic of the graded-refractive-index film. The light enters the film from the glass side at an angle $\thetai$ and, depending on this angle, either is transmitted into the $\nb$ side at an angle $\thetat$ or suffers total internal reflection at some point and exits on the glass side. (b) Refractive-index profile of the system. (c) $R$ for $s$ and $p$ polarisations with $\tau=10\lambda$ and $\alpha=10^{-3}$. The qualitative features are the same for both polarisations.}\label{fig-layers}\end{figure}

We consider here the case of light with vacuum wave number $k_0$ travelling through a medium with refractive index $\ng=1.518$ (the refractive index of glass) until it reaches a medium with refractive index \smash{$n(z)=\nb+(\nw-\nb)\exp{-(z-z_0)/\tau\vphantom{a^2}}$} (exponential) or \smash{$n(z)=\nb+(\nw-\nb)\exp{-(z-z_0)^2/8\tau^2}$} (gaussian) with $\nb=1.333$ (the refractive index of water) and $\nw=1.35+\alpha i$; $z_0$ is the position where the refractive index changes from $\ng$ to $\nw$, $\tau$ is the penetration depth of the gradient, and $\nw$ is the ``wall refractive index''. The boundary between the media is perpendicular to the $z$ direction, and the light is incident on the first boundary at an angle $\thetai$. We model the region where $n(z)$ changes appreciably, which we shall henceforth call the ``graded-refractive-index film'' (GRIF), as a collection of $N$ parallel layers of equal thickness $\Delta=D/N$ and refractive index \smash{$n_\ell=\nb+(\nw-\nb)\exp{-\ell\Delta/\tau}$} or \smash{$n_\ell=\nb+(\nw-\nb)\exp{-\ell^2 \Delta^2/8\tau^2}$} ($\ell\in\{0,\ldots,N-1\}$), accordingly (figure~\ref{fig-layers}a and~\ref{fig-layers}b), and take the limit as $N\rightarrow\infty$. In both cases, taking the limit $\tau\rightarrow0$ or the limit $\tau\rightarrow\infty$ equates to having a simple boundary between two regions with refractive indices $\ng$ and either $\nb$ ($\tau\rightarrow0$) or $\nw$ ($\tau\rightarrow\infty$).

We first employ the transfer-matrix method \cite{ref-Born-POO} to calculate the Fresnel reflection and transmission coefficients at each boundary and ultimately calculate the reflectance of the system. Briefly, the transfer matrix for the $\ell$-th layer of the GRIF is
\eq{M_\ell & = & \frac{e^{-ik_z^{(\ell)}\Delta}}{t_{\ell,\ell+1}}\,\begin{pmatrix}1 & r_{\ell,\ell+1}\\r_{\ell,\ell+1}e^{2ik_z^{(\ell)}\Delta} & e^{2ik_z^{(\ell)}\Delta}\end{pmatrix},}
where $k_z^{(\ell)}$ is the $z$ component of the wave vector in the $\ell$-th layer, $r_{\ell,\ell+1}$ and $t_{\ell,\ell+1}$ are the Fresnel reflection and transmission coefficients at the boundary between the $\ell$-th and $(\ell+1)$-th layers, and the $(N+1)$-th layer is considered to be the ``bulk'' medium with constant refractive index $\nb$. The total transfer matrix of the GRIF is $\Mg M_1 M_2\cdots M_{N-2}M_{N-1}$, where $\Mg$ is the transfer matrix for the boundary between the glass (i.e.~$\ell=0$) and the region with $\ell=1$ (for this matrix we take $\Delta=0$). If we write the total matrix as
\eq{M & = & \begin{pmatrix}m_{1,1} & m_{1,2}\\m_{2,1} & m_{2,2}\end{pmatrix},}
then the reflection coefficient of the GRIF is
\eq{r & = & \frac{m_{2,1}}{m_{1,1}}.\label{eq-r}}
The reflectance is then given by $R=|r|^2$.

\section{Discussion and conclusions}\label{sec-R}

In figure~\ref{fig-layers}c we can see that the influence of the refractive-index gradient on the reflectance ($R$) curves is noticeable only in a small angle interval near the BR critical angle $\thetab$ and between $\thetab$ and the WR critical angle $\thetaw$. Outside this small interval, the curves are indistinguishable from the reflectivity of a simple interface. The qualitative features of the curves, as well as all quantitative features except the amplitude of the peaks, are the same for both polarisations. Thus, we will henceforth only concern ourselves with the $s$ polarisation.

\begin{figure}[b!]\begin{center}\begin{subfigure}[t]{\figwidthc}\begin{center}{\includegraphics[width=\figwidthc]{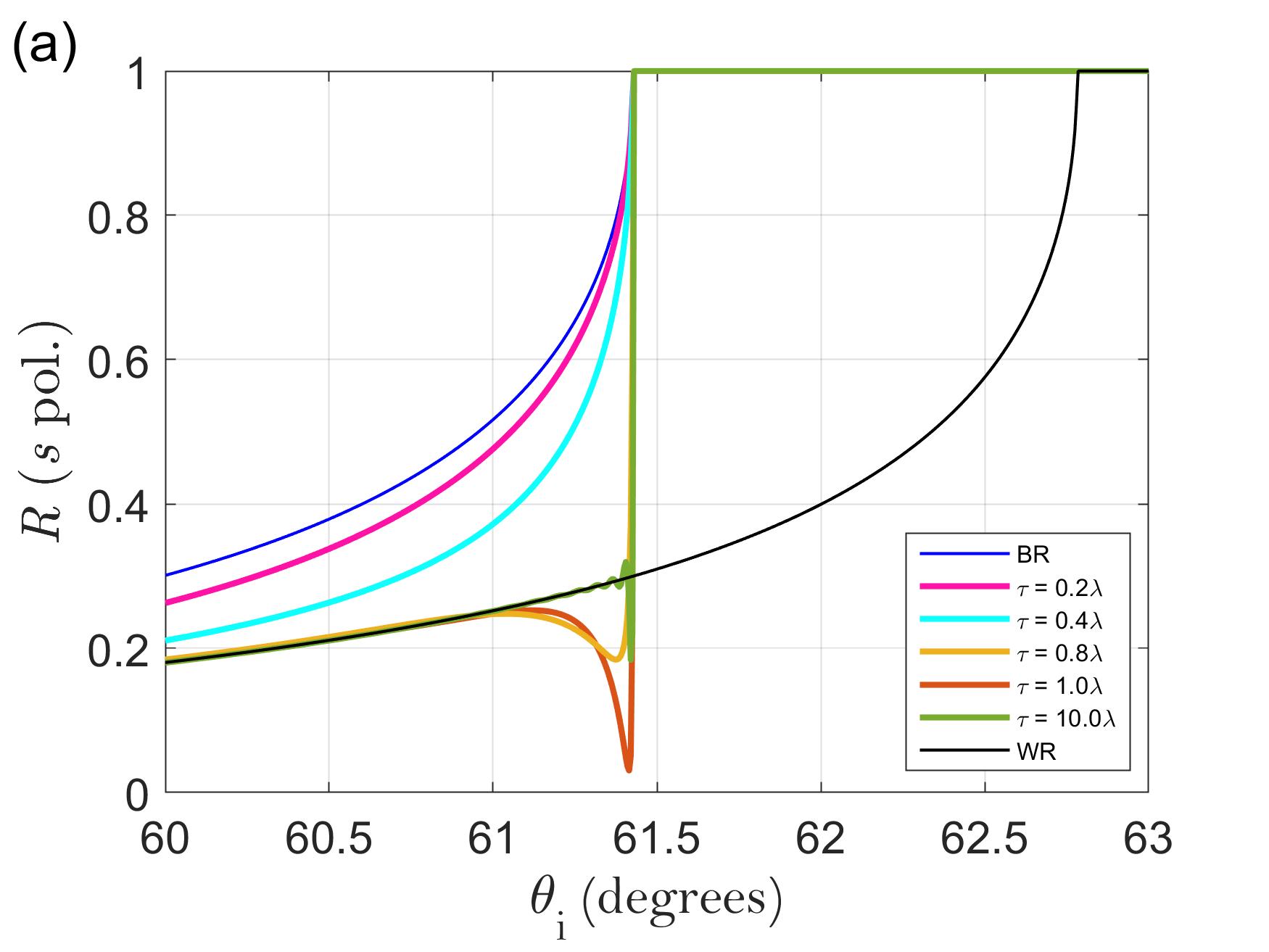}}\end{center}\end{subfigure}\\\begin{subfigure}[t]{\figwidthc}\begin{center}{\includegraphics[width=\figwidthc]{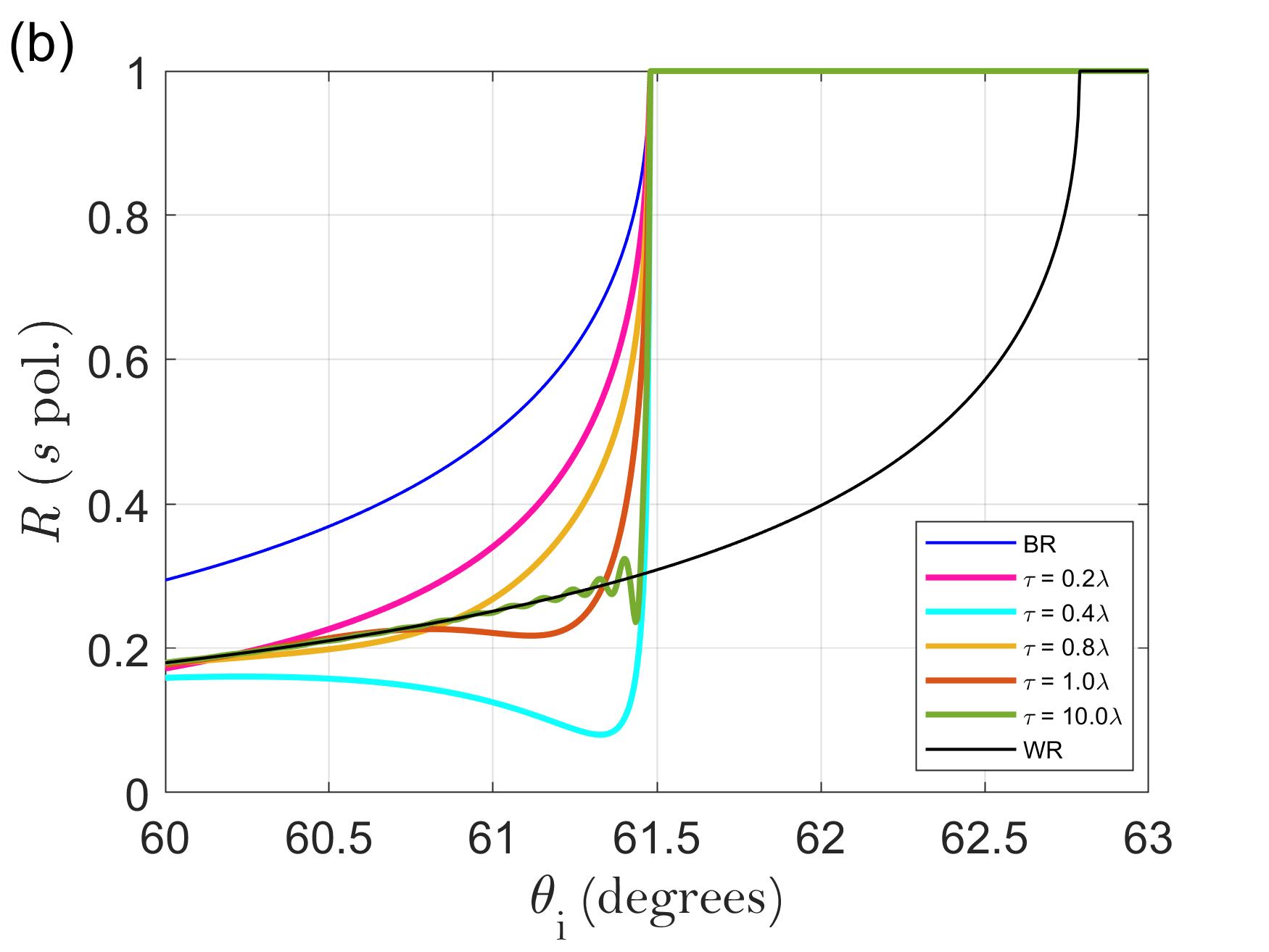}}\end{center}\end{subfigure}\captionsetup{singlelinecheck=off}\end{center}\caption[.]{Dependence of the reflectance on $\tau$ for $s$ polarisation in the exponential (a) and gaussian (b) cases. Note that the oscillations occur only for values of $\tau$ from about $0.8\lambda$ (exponential profile) or $0.4\lambda$ (gaussian profile).}\label{fig-tau}\end{figure}

Figure~\ref{fig-tau} shows the reflectance of the system as a function of the angle of $\thetai$ for $s$ polarisation for different values of $\tau$ and with $\alpha=0$. The black and blue curves correspond to the reflectance of a simple boundary between two media corresponding to the limit cases mentioned before ($\tau\rightarrow0$ and $\tau\rightarrow\infty$); for brevity, we call these the bulk reflectance (BR) and the wall reflectance (WR), respectively. In all cases with finite $\tau$, $R$ reaches 1 at $\thetab$, which is obvious given that for $\thetai\geq\thetab$ there will be TIR at some point in the collection of layers and all the light will be reflected back into the glass.

Several things are noteworthy. First, the higher $\tau$ is the more closely $R$ follows the WR up to $\thetab$, where it jumps to 1 due to TIR with the bulk. Second, there is a threshold value of $\tau$ above which oscillations prior to $\thetab$ become visible; this value is between $0.4\lambda$ and $0.8\lambda$ for the exponential profile and between $0.2\lambda$ and $0.4\lambda$ for the gaussian profile, which can be explained as follows: a refractive index with a gaussian profile falls more slowly than one with an exponential profile, which means a lower value of $\tau$ still shows a gradient in the gaussian case, whereas for the exponential case very small values of $\tau$ imply too rapid a fall and the gradient becomes indistinguishable from an immediate fall from $\ng$ to $\nb$ (see, for example, the $\tau=0.4\lambda$ and $\tau=0.2\lambda$ curves in figure~\ref{fig-tau}a, which are very similar to the BR curve). This leads us to our third remark, which is that, for $\tau$ above the aforementioned threshold, reflectance measurements for $\thetai<\thetab$ will reveal the presence of a gradient (as opposed to a simple interface) by the existence of oscillations in the reflectance followed by a sudden increase at $\thetab$ (as opposed to a gradual increase). Finally, for very large $\tau$ the oscillations before $\thetab$ become smaller and eventually disappear because the gradient becomes very slow; we can see the beginning of this effect in the $\tau=10\lambda$ curves.

\begin{figure}[t!]\begin{center}\begin{subfigure}[t]{\figwidthc}\begin{center}{\includegraphics[width=\figwidthc]{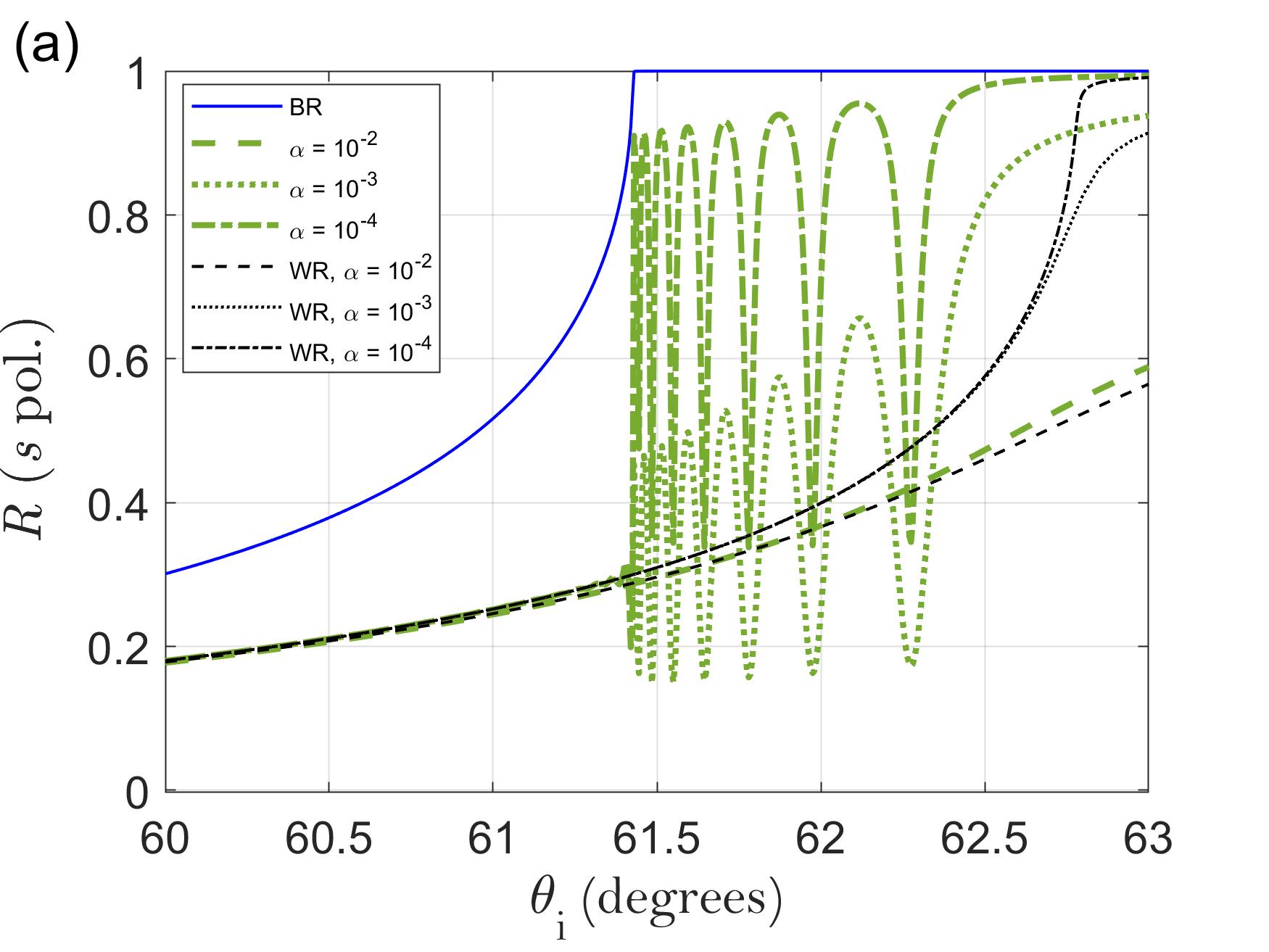}}\end{center}\end{subfigure}\\\begin{subfigure}[t]{\figwidthc}\begin{center}{\includegraphics[width=\figwidthc]{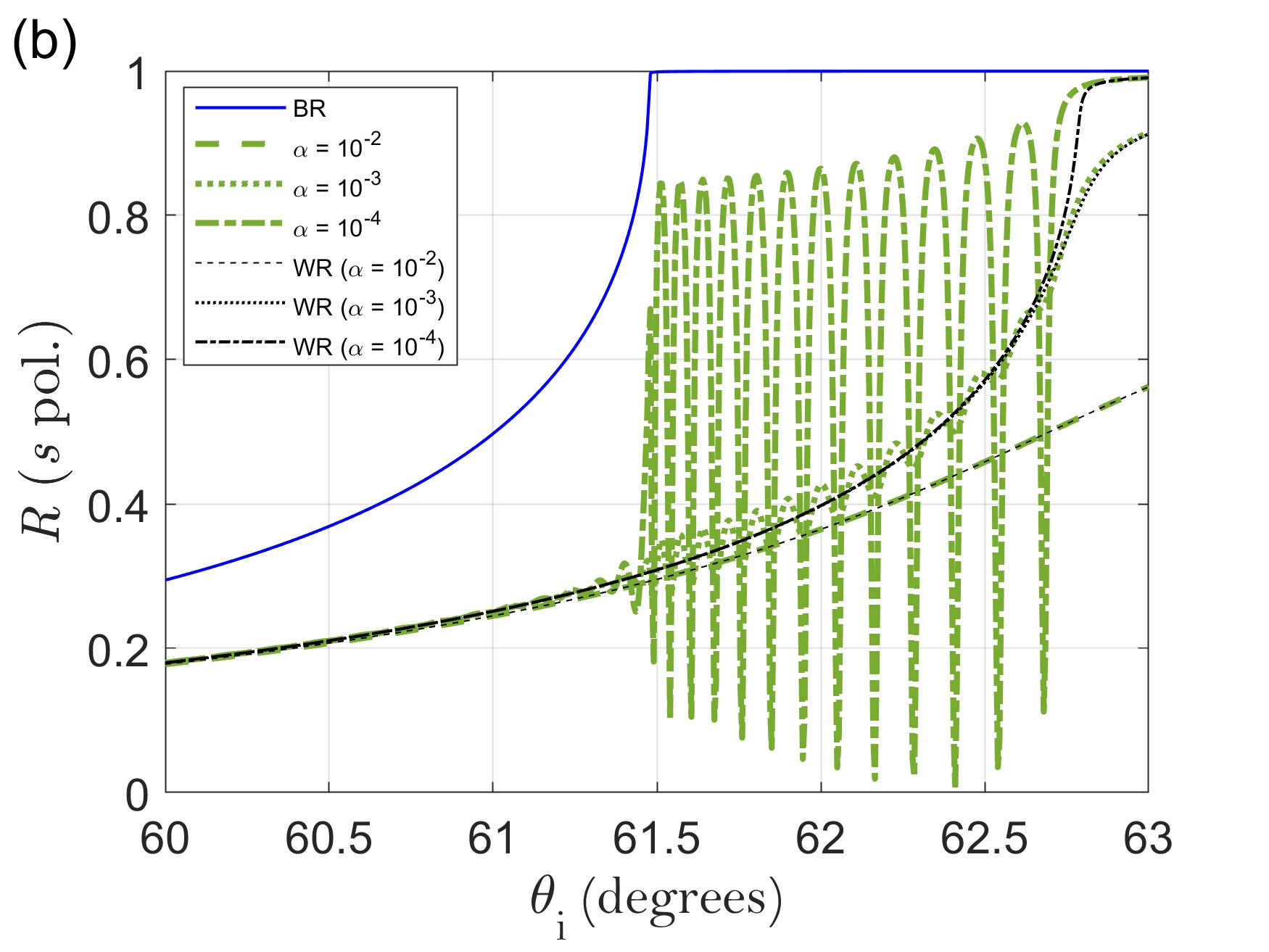}}\end{center}\end{subfigure}\captionsetup{singlelinecheck=off}\end{center}\caption[.]{(a) Dependence of $R$ on $\alpha$ in the exponential case for $s$ polarisation ($\tau=10\lambda$ was used. The peaks with $\alpha=10^{-4}$ and $\alpha=10^{-3}$ are caused by interference (see main text); the extinction with $\alpha=10^{-2}$ is too high for this to occur. (b) As (a) for the gaussian case.\vspace{-0.2in}}\label{fig-alpha}\end{figure}

\begin{figure}[b!]\begin{center}\begin{subfigure}[t]{\figwidthc}\begin{center}{\includegraphics[width=\figwidthc]{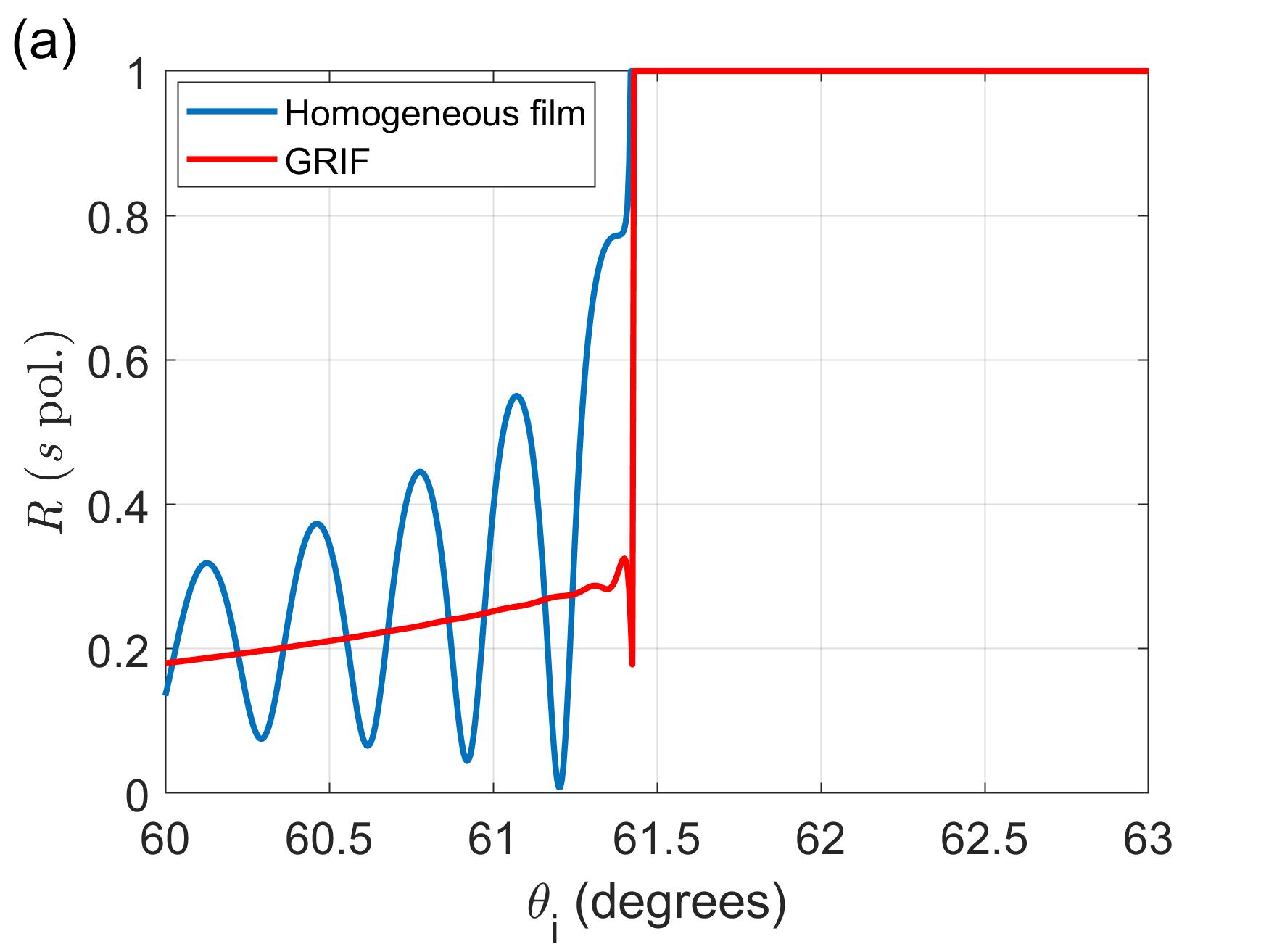}}\end{center}\end{subfigure}\\\begin{subfigure}[t]{\figwidthc}\begin{center}{\includegraphics[width=\figwidthc]{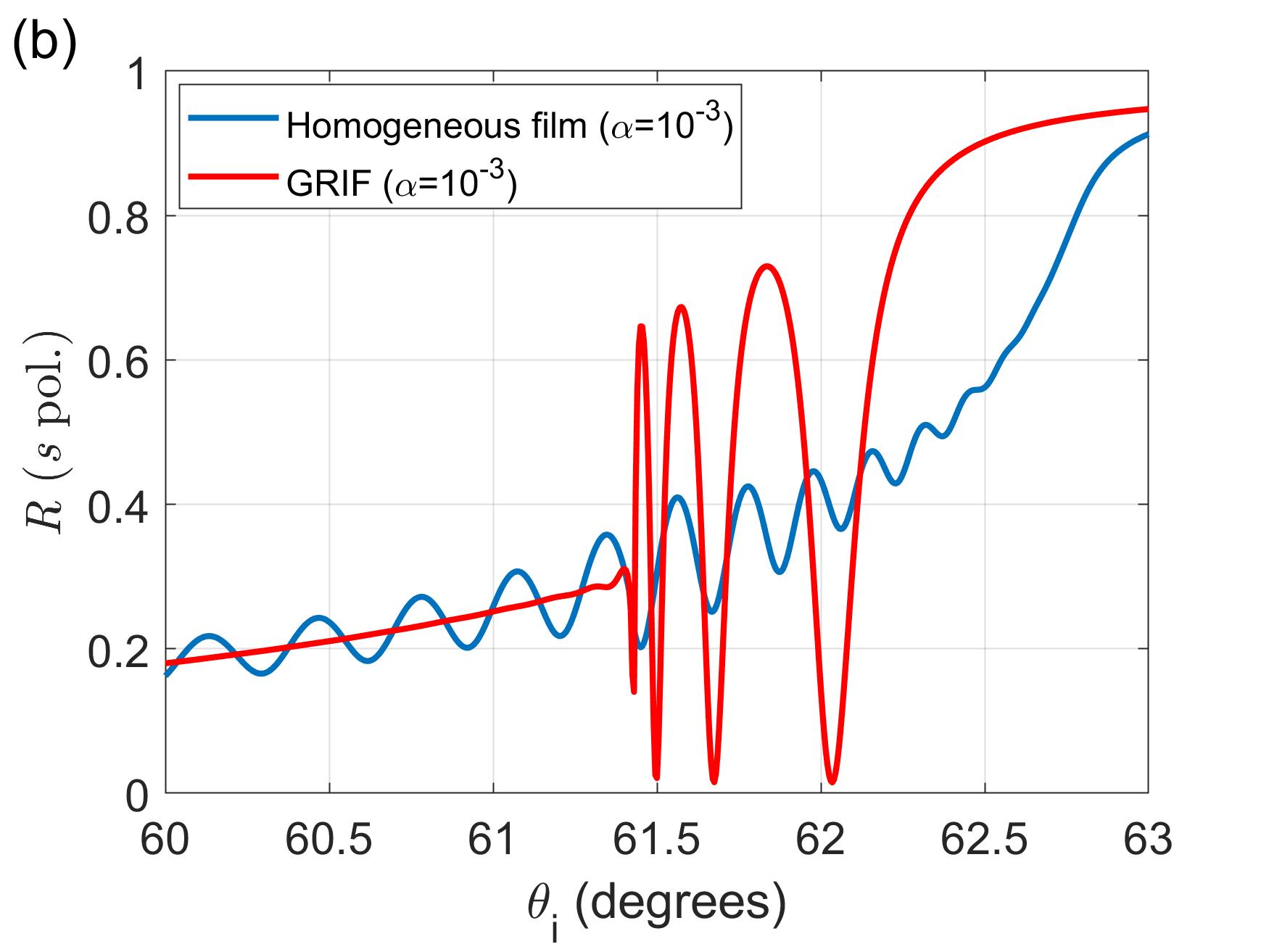}}\end{center}\end{subfigure}\captionsetup{singlelinecheck=off}\end{center}\caption[.]{Comparison of the reflectance of a homogeneous film with refractive index $\nw$ and that of a GRIF with $\alpha=0$ (a) and $\alpha=10^{-3}$ (b). In both cases, $\tau=5\lambda$, the GRIF refractive index has an exponential profile, and the thickness of the homogeneous film is the same as that of the GRIF.}\label{fig-uniform}\end{figure}

Figure~\ref{fig-alpha} shows $R$ for $\tau=10\lambda$ and various values of $\alpha$ between $10^{-4}$ and $10^{-2}$. Several important conclusions can be derived from these curves. Because all of the interesting features of $R$ occur between $\thetab$ and $\thetaw$, the remainder of our analysis will focus on this region. Note that, because $\nw$ now has a non-zero imaginary part, TIR is not achieved immediately after $\thetaw$ due to absorption.

The oscillations between $\thetab$ and $\thetaw$ are the effect of interference of the different reflection orders. In the case $\alpha=0$ (figure~\ref{fig-tau}), since there is no extinction, all light eventually returns to the glass (assuming the media are infinite in the $x$ and $y$ directions), so there are no oscillations. With extinction, however, eventually there is no more reflected light to compensate the destructive interference of the first few reflection orders and the effects of this interference can be seen. However, once the extinction is high enough (which occurs somewhere between $\alpha=10^{-3}$ and $\alpha=10^{-2}$), not even the second reflection order can make it back and the oscillations are lost; here $R$ simply follows the WR.

The number of oscillations between $\thetab$ and $\thetaw$ is roughly equal to $0.75\tau/\lambda$ for the exponential profile and $1.3\tau/\lambda$ for the gaussian profile (rounded to the nearest integer; data not shown). The fact that the number (and the positions) of the oscillations depends on $\tau/\lambda$, which is effectively a measure of the thickness of each homogeneous layer relative to the wavelength, as well as of the contrast between successive layers, indicates that the oscillations are caused by interference, as does the fact that neither the number nor the positions are affected by $\alpha$ (as long as it is non-zero but small enough to allow said interference to take place).

In all cases, the angle frequency of the oscillations (that is, how many oscillations per unit $\thetai$ occur) becomes gradually lower as $\thetai$ increases. This is becuase, as the angle of incidence increases, the optical path length inside each layer also increases, which means that the light spends longer in an absorbing layer before reaching the next layer; ultimately, this leads to the effective thickness of the GRIF (i.e.~the thickness the light is able to travel through before becoming too strongly attenuated) decreasing with increasing $\thetai$. Compare this to a homogeneous thin film, where the angle frequency of the oscillations is also a decreasing function of the film thickness.

It is also noteworthy that all the curves with $\alpha\neq0$ exhibit a slightly smaller apparent wall critical angle (manifested as the shoulder after which no more oscillations occur) than the corresponding WR curves. Recall that the WR case corresponds to a simple interface between glass and an infinite medium with constant refractive index $\nw$. We believe this is again due to the effective thickness of the GRIF being smaller than its actual thickness; in this case, after the last reflectance oscillation, the light that enters the GRIF interacts with only a very thin section at the top of it before being reflected with minimal attenuation, so the refractive index the light ``sees'' is, in fact, not $\nw$, but an index whose real part is slightly smaller than $\nw$, which leads to a reduced apparent critical angle.

For gradients with smaller values of $\Real{\nw}-\nb$, our results will not change except for the angle region between $\thetab$ and $\thetaw$ becoming narrower; likewise, if this quantity increases, said region will become wider. If $\nw$ is real, a smaller $\nw-\nb$ will make the $R$ curve before $\thetab$ resemble the BR curve more closely and the jump at $\thetab$ will become smaller; for small enough values of $\nw-\nb$, this jump will become difficult to detect.

It is worthwhile to compare the reflectance of a GRIF with that of a film with constant refractive index $\nw$ (figure~\ref{fig-uniform}). Such a comparison reveals the means by which one may distinguish an optical gradient from a uniform thin film. In the case of real $\nw$, they are only distinguishable at $\thetai<\thetab$, since the reflectance equals 1 for all angles after the bulk critical angle. The main difference is the amplitude of the oscillations of $R$, which are much larger for a homogeneous film. In the case of complex $\nw$, the previous difference is still present, but additional differences appear after $\thetab$: the oscillations become larger and have variable angle frequency for a GRIF, while they have constant angle frequency for a homogeneous film.

Hence, not only is it possible to detect a gradient by angle-resolved reflectometry both before and after the bulk critical angle, but it is sometimes possible to characterise it. By now, it should be clear how to tell when $\nw$ is complex. As shown previously, there are differences between a gaussian profile and an exponential profile; there will, of course, also be differences between these two profiles and others, so one must have in mind the types of profiles one would expect from the system under consideration, depending on the nature of the physical processes involved. Once the type of refractive-index gradient has been identified, however, the number of oscillations between $\nb$ and $\nw$ (provided $\nw$ is complex but does not have too large an imaginary part) provides insight into the steepness of the gradient.

We conclude by restating the fact that it is possible to detect and characterise a refractive-index gradient by $R$ measurements just before $\thetab$ and between $\thetab$ and $\thetaw$ assuming the gradient is neither too fast ($\tau\lesssim0.2\lambda$) nor too slow ($\tau\gtrsim10\lambda$). In the case of complex $\nw$, the number of interference peaks between $\thetab$ and $\thetaw$ are directly a function of the type of gradient and how fast it is, which allows us to obtain estimates for these parameters. If the imaginary part of $\nw$ is too high ($\gtrsim10^{-2}$), however, these peaks disappear and the gradient becomes undetectable. For reasonably absorbing gradients, however, $R$ measurements can be used to detect and characterise processes involving flow, sedimentation, diffusion, ablation, biochemical interaction and temperature gradients.

\section{References}

\begingroup
\renewcommand{\section}[2]{}

\endgroup

\vfill

\end{document}